\pdfoutput=1
\documentclass[preprint,12pt,double]{elsarticle}
\usepackage[nomarkers,nofiglist, notablist]{endfloat}
\usepackage{lineno}
\usepackage{subcaption}
\usepackage{multicol}
\usepackage{setspace}
\usepackage{float}
\usepackage{amsmath,amssymb,amstext,bm,array,multirow,booktabs} 
\usepackage{graphicx} 
\graphicspath{{figures/}}
\usepackage{bibentry}

\newcommand{\ignore}[1]{}
\newcommand{\nobibentry}[1]{{\let\nocite\ignore\bibentry{#1}}}

\modulolinenumbers[5]

\begin{document}
\nobibliography*
\title{Atomistic Kinetic Monte Carlo Simulations of Polycrystalline Copper Electrodeposition}

\author[UW]{Tanyakarn Treeratanaphitak}
\author[UW]{Mark D. Pritzker}
\ead{pritzker@uwaterloo.ca}
\author[UW,WIN]{Nasser Mohieddin Abukhdeir\corref{cor}}
\cortext[cor]{Corresponding author. Phone: +1 (519) 888-4567 x31306}
\ead[url]{http://chemeng.uwaterloo.ca/abukhdeir/}
\ead{nmabukhdeir@uwaterloo.ca}

\address[UW]{Department of Chemical Engineering}
\address[WIN]{Waterloo Institute for Nanotechnology\\University of Waterloo\\200 University Avenue West\\Waterloo, Ontario, Canada N2L 3G1}

\biboptions{sort&compress}
\journal{Electrochemistry Communications}

\begin{frontmatter}

\begin{abstract}
A high-fidelity kinetic Monte Carlo (KMC) simulation method (\nobibentry{Treeratanaphitak2014}) using the semi-empirical multi-body embedded-atom method (EAM) potential has been extended to model polycrystalline metal electrodeposition.
Simulations using KMC-EAM are performed over a range of overpotentials to predict the effect on deposit texture evolution.
Roughness-time power law behaviour ($\propto t^\beta$) is observed where $\beta=0.62 \pm 0.12$, which is in good agreement with past experimental results.
Furthermore, the simulations provide insights into the dynamics of sub-surface deposit morphology which are not directly accessible from experimental measurements.
\end{abstract}


\end{frontmatter}

\linenumbers

\section{Introduction}

Electrodeposition is widely used to fabricate micro- and nano-structures for various applications including interconnects \cite{Andricacos1998}, electrodes \cite{Elias2007}, catalysts and sensors \cite{Zhang2012}.
The preferred deposit morphology that yields optimal performance varies depending on the application.
Thus, it is important to better understand the processes occurring during electrodeposition and the relationships between process parameters and growth kinetics.

Atomistic simulations of electrodeposition can help predict the relationship between process parameters and kinetics.
These simulations can be used to enhance and focus experimental research through identification of key processes and parameter regimes.
Kinetic Monte Carlo (KMC) methods are an important class of simulation methods capable of modelling the dynamics of the evolution of electrodeposition at experimentally-relevant time scales.
Recent advances \cite{Treeratanaphitak2014,Gimenez2003,Huang2009} in high-fidelity atomistic KMC simulations using the highly descriptive embedded-atom method (EAM) potential \cite{Daw1984} have moved the state-of-the-art closer to being able to make direct comparisons with experimental data from electrodeposition processes.

The traditional approach to KMC simulation of polycrystalline deposition is to use the so-called ``1+1'' dimensional solid-on-solid model (SOS) \cite{Bruschi2000,Liu2009EC,Ruan2010,Liu2013}.
The SOS approach, while computationally efficient, has several limitations when comparing simulated morphologies to atomic-resolution deposits and in the severe approximations made regarding deposit energetics \cite{Treeratanaphitak2014}.
Instead of computing the energy of the deposit from an interaction potential, the SOS method treats the energy at each site as a sum of two terms that scale linearly as a function of coordination number.

In a recent advance, Huang et al. \cite{Huang2009} simulated two-dimensional nickel electrodeposition under kinetically-controlled conditions in the presence of hydrogen atoms.
This method used a high-fidelity atomistic resolution of the deposit and the EAM potential for Ni-Ni and Ni-H interactions \cite{Daw1984}.
Polycrystalline simulation using KMC and the EAM potential was also performed for vapour deposition by Gilmer et al \cite{Huang1998,Zepeda-Ruiz2010} by assigning orientation angles to sites, but restricting their positions to a single lattice.
The deposit energy was evaluated using only first nearest neighbours, which does not completely describe the interaction energy of the atoms using EAM.
Rubio et al. \cite{Rubio2003} extended the method to represent the polycrystalline structure with multiple lattices, but retained the first nearest neighbour assumption.
This restriction on the interaction potential can influence deposit morphologies obtained from simulations.
These polycrystalline EAM/KMC methods were not developed for electrodeposition and do not include terms for deposition/dissolution kinetics.

In this work, a KMC method is presented and used to simulate polycrystalline electrodeposition using the multi-body EAM potential (KMC-EAM) which includes collective diffusion mechanisms, deposition/dissolution mechanisms and direct resolution of atomic polycrystalline morphologies.
The KMC-EAM method is applied to potentiostatic copper electrodeposition onto an atomically smooth polycrystalline copper substrate.
Spatially varying deposition and dissolution rates are resolved from the difference in surface energies of the crystal faces that are exposed.
Simulations are performed to model potentiostatic deposition over a range of overpotentials to determine the resulting evolution of roughness and deposition rate.

\section{Model Description}

The presented polycrystalline KMC-EAM method is an extension of the single-crystal KMC-EAM method from ref. \cite{Treeratanaphitak2014} and is applied to simulate potentiostatic deposition of copper.
The KMC-EAM method is a three-dimensional kinetic Monte Carlo method that uses a direct atomistic representation of a metal deposit.
The total energy or Hamiltonian contains the highly descriptive EAM interaction potential which is fit to quantum mechanical simulations, the experimentally obtained lattice constant, elastic constant and sublimation energy \cite{Daw1984}.
The specific EAM parameters used for copper are obtained from ref. \cite{Adams1989}.
Another unique aspect of the KMC-EAM method is that it includes collective diffusion mechanisms such as atom exchange and step-edge atom exchange, which have been shown to be vital to accurately simulate single-crystal deposition morphology \cite{Treeratanaphitak2014}.

The extension of KMC-EAM to polycrystalline electrodeposition requires modelling atomic configurations where metal atoms can occupy different crystal lattices.
Since the EAM potential does not require that atoms reside on a fixed lattice, evaluation of the energy of the system requires no modifications to the potential.
Each grain is approximated to reside on a different lattice of arbitrary (user-selected) orientation, which is set as an initial condition.
During simulation, electrodeposition occurs only on grain surfaces that are exposed while collective diffusion mechanisms are permitted on all grain surfaces.
When grain boundaries are in close proximity to each other, mutual grain growth on unoccupied sites within an impingement distance of $a\sqrt{2}/2$ is inhibited, where $a$ is the lattice spacing.

The propensities (or rates) of deposition and dissolution for each possible site are governed by the current density $i$ \cite{Budevski1996}:
\begin{align}
\Gamma_{dep} &= \frac{i}{z~e~n_{dep}} = \frac{i_{Cu}^{0}}{z~e~n_{dep}}\exp\left(-\frac{\alpha_{c}\eta}{k_{B}T}\right),
\label{eq:dep_rate_old} \\
\Gamma_{diss} &= \frac{i}{z~e~n_{diss}} = \frac{i_{Cu}^{0}}{z~e~n_{diss}}\exp\left(\frac{\alpha_{a}\eta}{k_{B}T}\right),
\label{eq:diss_rate_old}
\end{align}
where the current density is calculated from the Butler-Volmer equation at the specified overpotential using parameters from ref. \cite{Caban1977}, $e$ is the elementary charge, $z$ is the number of electrons transferred in the reduction reaction, $i_{Cu}^{0}$ is the exchange current density, $\alpha_{a}$/$\alpha_{c}$ is the anodic/cathodic transfer coefficient, $\eta$ is the overpotential and $n_{dep}$/$n_{diss}$ is the number of sites per surface area ($\mathrm{m}^{-2}$) available for deposition/dissolution.
The rates of deposition and dissolution are not uniform across the surface and depend on the surface energy of the exposed grain crystal face.
The surface energy computed from the embedded-atom method does not take into account interactions with the electrolyte and, as a result, approximations which conform to the macroscopic deposition/dissolution rates are used.
These approximations involve linearization of the propensity about the average change in energy resulting from deposition/dissolution computed from the EAM potential; the resulting propensities at each possible site are:
\begin{align}
\Gamma_{i,dep}' &= \Gamma_{dep}  \frac{\Delta^\downarrow E_{i}}{\Delta^\downarrow E_{avg}}, \label{eq:dep_rate} \\
\Gamma_{j,diss}' &= \Gamma_{diss} \left(2 - \frac{\Delta^\uparrow E_{j}}{\Delta^\uparrow E_{avg}}\right), \label{eq:diss_rate} 
\end{align}
where $\Delta^\downarrow E_{avg}$/$\Delta^\uparrow E_{avg}$ is the average energy difference over all possible deposition/dissolution events and $\Delta^{\downarrow} E_{i}$/$\Delta^{\uparrow} E_{j}$ is the difference in energy resulting from deposition/dissolution at site $i$/$j$.

Propensities of diffusion events follow an Arrhenius-type relationship \cite{Treeratanaphitak2014}:
\begin{equation}
\Gamma_{i,d} = \begin{cases}
                \nu_{d} \exp\left(-\frac{E_{d}}{k_{B}T}\right) & \Delta E \leq 0 \\
                \nu_{d} \exp\left(-\frac{E_{d}+\Delta E}{k_{B}T}\right) & \Delta E > 0
                \end{cases}
\label{eq:diff_rate}
\end{equation}
where $\nu_{d}$ is the atomic vibrational frequency and $E_{d}$ is the activation energy of one of the following diffusion events: hopping, atom exchange, step-edge atom exchange or grain boundary migration/diffusion.
Since the determination of the activation energies of diffusion using the EAM potential \cite{Antczak2010} is computationally prohibitive, previously computed values are used instead.
The activation energy of grain boundary diffusion is assumed to be $0.5~\mathrm{eV}$ while the other activation energies are the same as those used in the previous single-crystal study \cite{Treeratanaphitak2014}: $E_{hopping} = 0.5~\mathrm{eV}$, $E_{step} = 0.2~\mathrm{eV}$ and $E_{exch} = 0.7~\mathrm{eV}$ \cite{Antczak2010}.

Initial conditions are used to replicate electrodeposition onto an atomically smooth polycrystalline copper (FCC) substrate which enables nucleation to be neglected.
These initial substrates are established using randomly generated ``seed'' lattices with surface orientations of (100), (111) and (110).
Periodic boundary conditions are used at the $x$- and $y$-boundaries to approximate bulk surface deposition.
Copper reduction is assumed to proceed by a one-step reaction under conditions where it is kinetically controlled, i.e., mass transfer from the electrolyte plays no role.
All simulations are performed for deposition of a given number of atoms ($7\times 10^{4}$) within a simulation domain of $3023~\mathrm{nm}^{3}$ ($14.46~\mathrm{nm} \times 14.46~\mathrm{nm} \times 14.46~\mathrm{nm}$) at different overpotentials.
The required computational time for the simulations ranged from 2--7 days using single CPU core serial processes.
The KMC-EAM implementation used in this work has been made freely available \footnote{\url{http://launchpad.net/mckmc}}.

\section{Results and Discussion}

\begin{figure*}
    \centering
    \includegraphics[width=\linewidth]{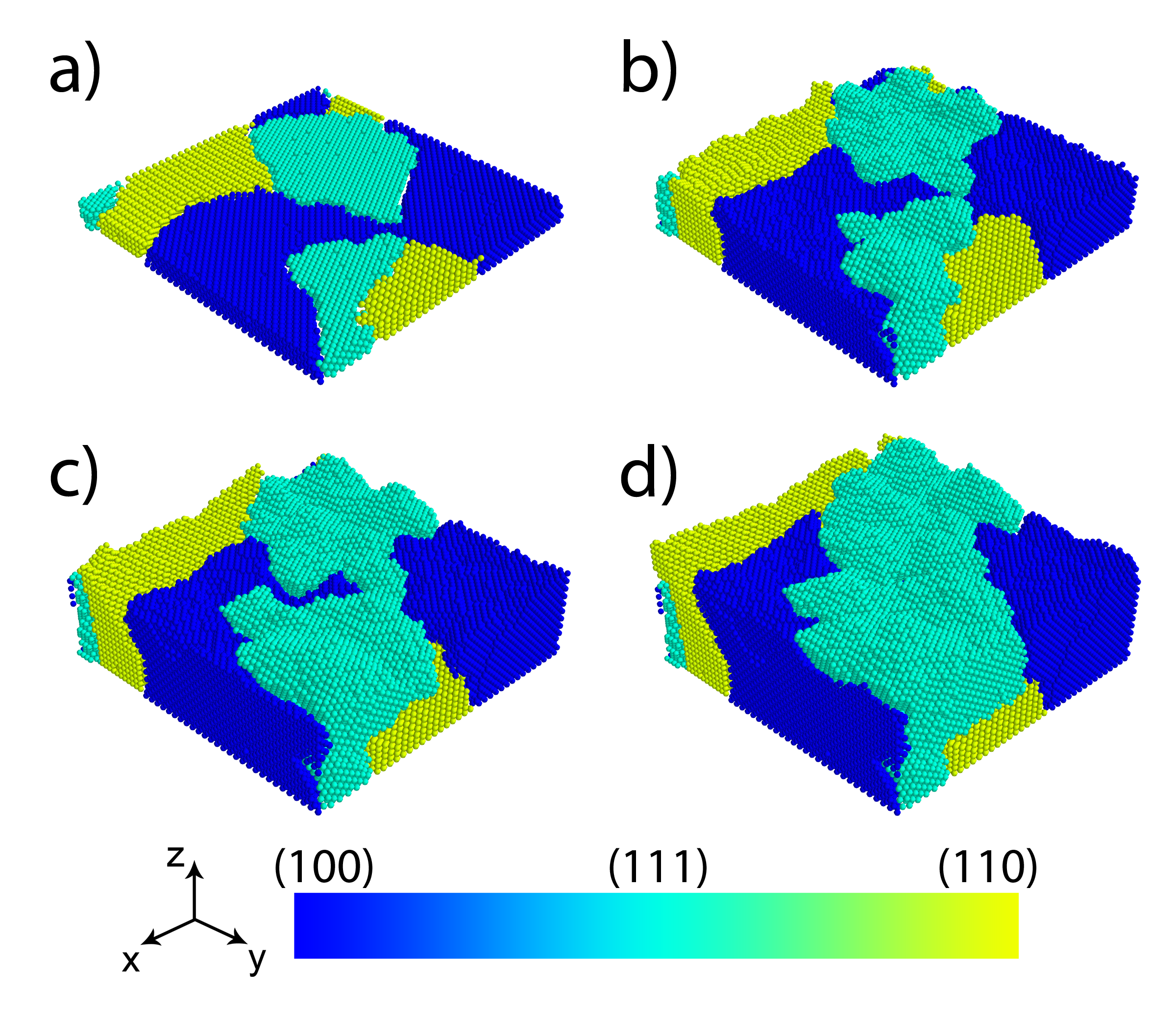}
    \caption{Simulated evolution of an electrodeposit after a) 0\% ($t=0~\mathrm{s}$), b) 33.3\% ($t=0.05~\mathrm{s}$), c) 66.7\% ($t=0.09~\mathrm{s}$) and d) 100\% ($t=0.13~\mathrm{s}$) of the $7 \times 10^4$ atoms have been deposited at $\eta = -0.15~\mathrm{V}$. Simulation domains are $14.46~\mathrm{nm} \times 14.46~\mathrm{nm} \times 14.46~\mathrm{nm}$.}
    \label{fig:sample_domains}
\end{figure*}

\begin{figure*}
    \centering
    \includegraphics[width=\linewidth]{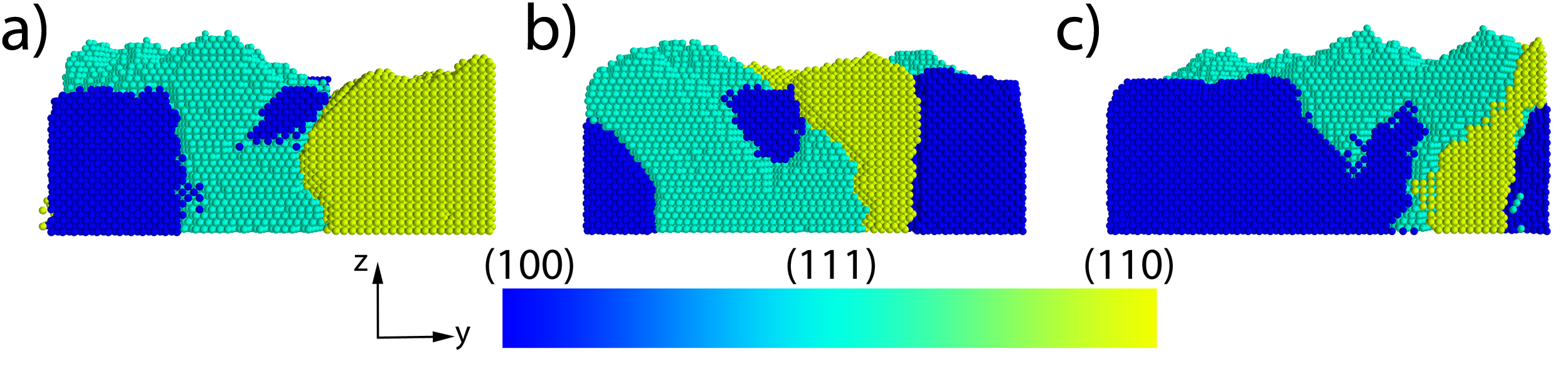}
    \caption{Cross-section view of simulated deposit morphologies with $7 \times 10^{4}$ atoms deposited at a) $\eta = -0.05~\mathrm{V}$, b) $\eta = -0.10~\mathrm{V}$ and c) $\eta = -0.15~\mathrm{V}$. Simulation domains are $14.46~\mathrm{nm} \times 14.46~\mathrm{nm} \times 14.46~\mathrm{nm}$.}
    \label{fig:sideview}
\end{figure*}

Figures \ref{fig:sample_domains}a-d show the simulated evolution of a deposit from an initially smooth condition when $\eta = -0.15~\mathrm{V}$.
Figures \ref{fig:sample_domains}b-d exhibit the general trends observed in all simulations which reveal that the (111) surface grows at a faster rate than the other surfaces.
Additionally, the (111) grains exhibit three-dimensional growth while the (100) grains exhibit primarily layer-by-layer two-dimensional growth.
Figures \ref{fig:sideview}a-c show cross-sections of final deposit morphologies obtained from simulations with three different values of $\eta$.
These figures more clearly show the differences in grain growth among different surface orientations and that the (110) surface morphology is intermediate to that of the other two orientations.
These observed trends are in agreement with experimental observations of Cu/Cu(100) and Cu/Cu(111) homoepitaxy \cite{Ernst1992,Wulfhekel1996}.

\begin{figure*}
    \centering
    \includegraphics[width=\textwidth]{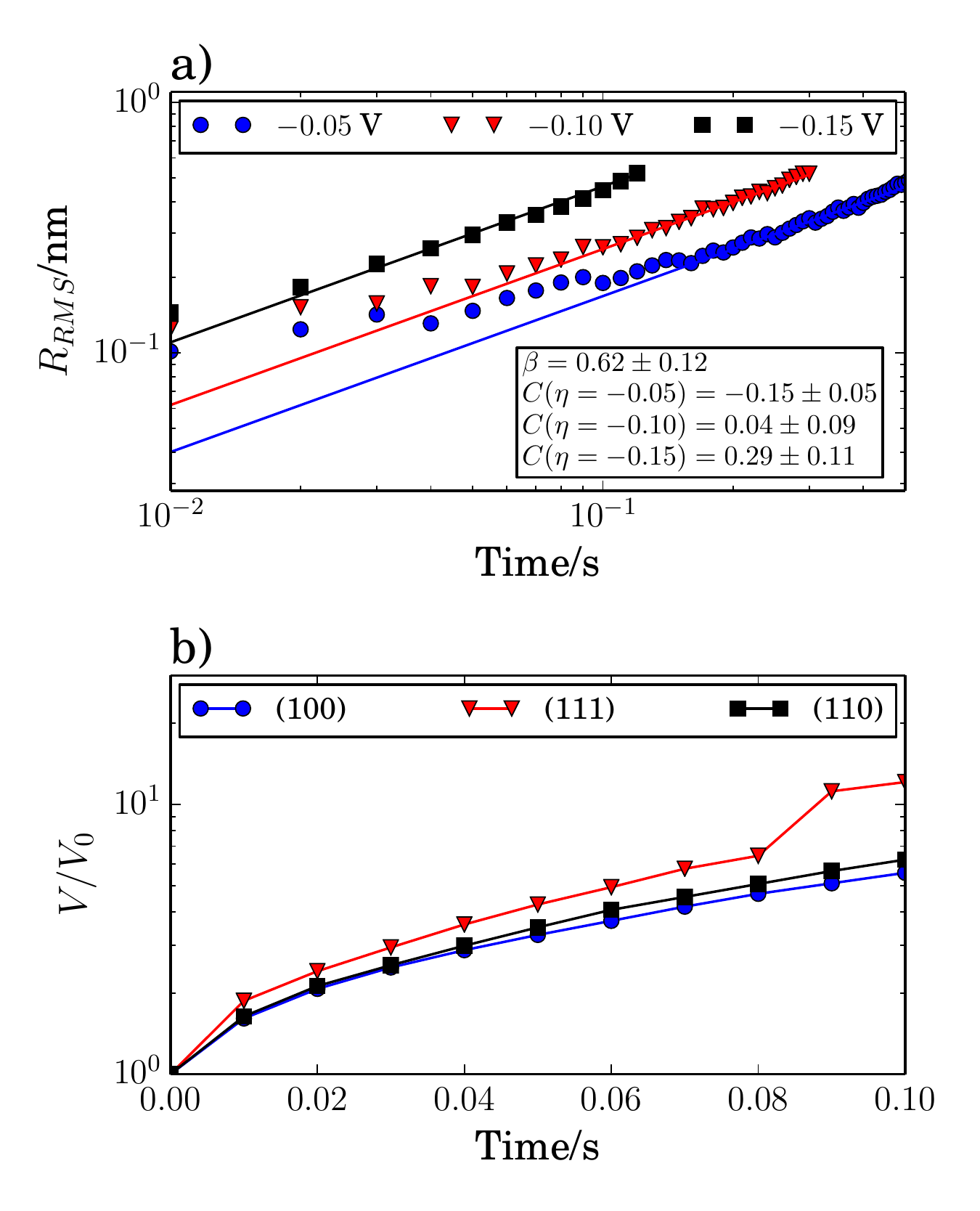}
    \caption{Evolution of a) RMS roughness versus time with power law fit $R_{RMS} = C(\eta)t^{\beta}$; b) ratio of grain volume over initial volume of the (100), (111) and (110) orientations with time at $\eta = -0.15~\mathrm{V}$. \label{fig:measurements}}
\end{figure*}

The evolution of the root-mean-squared roughness ($R_{RMS}$) is computed for sets of simulations with three different overpotentials (Figure \ref{fig:measurements}a). These simulations show that the roughness evolution obeys a power law relationship $R_{RMS} = Ct^{\beta}$ where $C$ depends on overpotential, while the exponent $\beta$ does not vary significantly with overpotential over the range considered and is found to be $0.62 \pm 0.12$. Experimental studies on metal electrodeposition in the absence of organic additives at current densities comparable to those considered in our simulations have shown that these systems often exhibit anomalous scaling whereby the surface roughness follows a power law relationship with time over distances shorter than a critical crossover length as well as over distances longer than this critical length \cite{Huo2001,Foster2005}. The spatial domain modelled in these simulations is not of sufficient size to carry out scaling analysis over length scales sampled in the experimental studies.

With one exception \cite{Schmidt1996}, experimental studies of roughness scaling on copper electrodeposition in acidic sulphate solutions in the absence of additives \cite{Iwamoto1994,Mendez1998,Schilardi1998,Huo2001,Drews2003,Osafo-Acquaah2006} have yielded $\beta$ values ranging from $0.8$ to $0.3$, including $0.63 \pm 0.08$ \cite{Mendez1998} and $0.60 \pm 0.05$ \cite{Schilardi1998} in close agreement with that obtained in our simulations. With regard to modelling results, a previous KMC simulation of this system using the standard SOS model obtained a $\beta$ value of only $0.04 \pm 0.06$ \cite{Drews2003}, considerably lower than that of the current study and  most of the experimental results. However, it must be acknowledged that the domain size and duration of electrodeposition considered in our simulations are much smaller than those in the experimental studies and that $\beta$ is strongly affected by length scale, electrode potential, current density, electrolyte composition, etc.   

Finally, Figure \ref{fig:measurements}b shows the evolution of the sub-surface of the deposit, which is inaccessible using current experimental methods. From the results for $\eta = -0.15~V$, the total volume of the (111) grain grows faster than the other two surface orientations. This is supported by past findings that the (111) orientation has the lowest surface energy \cite{Vitos1998,Jia2009} which is accounted for in the KMC-EAM method through eqn. \ref{eq:dep_rate}.

\section{Conclusions}

The KMC-EAM methodology has been extended from single-crystal to polycrystalline electrodeposition.
This includes capturing arbitrarily many grain orientations and the inclusion of additional mechanisms including atom dissolution and grain boundary diffusion.
Electrodeposition simulations are performed at an atomistic level and show that the (111) orientation grows preferentially in agreement with experimental observations.
Finally, the presented KMC-EAM method is found to predict roughness-time power law behaviour in agreement with experimental studies.
These results support the use of the KMC-EAM method to simulate the evolution of surface and sub-surface deposit morphology.

\section*{Acknowledgments}

This research was supported by the Natural Sciences and Engineering Research Council (NSERC) of Canada and the facilities of the Shared Hierarchical Academic Research Computing Network (SHARCNET).

\bibliographystyle{elsarticle-num-mod}
\bibliography{deposition,computational}

\end{document}